\documentclass{article}
\usepackage[utf8]{inputenc}
\usepackage[T1]{fontenc}
\usepackage{times}
\usepackage{helvet}
\usepackage{courier}

\usepackage{amsmath}
\usepackage{amssymb}
\usepackage{amsfonts}

\usepackage{booktabs}
\usepackage{multirow}
\usepackage{makecell}
\usepackage{array}
\usepackage{tabularx}

\usepackage{graphicx}
\usepackage{xcolor}
\usepackage{caption}

\usepackage[linesnumbered,ruled,vlined]{algorithm2e}

\usepackage{enumitem}
\usepackage{tcolorbox}
\usepackage{float}
\usepackage{nicefrac}
\usepackage{microtype}
\usepackage{mathtools}

\usepackage{url}
\usepackage[colorlinks=true, linkcolor=blue, citecolor=blue, urlcolor=blue]{hyperref}

\usepackage{tikz}

\usepackage[numbers,square]{natbib}

\usepackage{arxiv}

\usepackage{amsthm}

\title{LiveGraph: Active-Structure Neural Re-ranking for Exercise Recommendation}

\author{
    Rong Fu \\
    Independent Researcher \\
    Corresponding author \and
    Zijian Zhang \\
    Independent Researcher \and
    Jiekai Wu \\
    Independent Researcher \and
    Kun Liu \\
    Independent Researcher \and
    Xianda Li \\
    Independent Researcher \and
    Haoyu Zhao \\
    Independent Researcher \and
    Yang Li \\
    Independent Researcher \and
    Yongtai Liu \\
    Independent Researcher \and
    Ziming Wang \\
    Independent Researcher \and
    Rui Lu \\
    Independent Researcher \and
    Simon Fong \\
    Independent Researcher
}

\renewcommand{\headeright}{}
\renewcommand{\undertitle}{}
\renewcommand{\shorttitle}{LiveGraph}

\hypersetup{
    pdftitle={LiveGraph: Active-Structure Neural Re-ranking for Exercise Recommendation},
    pdfsubject={Computer Science - Information Retrieval (cs.IR); Computer Science - Machine Learning (cs.LG)}, 
    pdfauthor={Rong Fu, Zijian Zhang, Jiekai Wu, Kun Liu, Xianda Li, Haoyu Zhao, Yang Li, Yongtai Liu, Ziming Wang, Rui Lu, Simon Fong},
    pdfkeywords={Exercise Recommendation, Graph Neural Networks, Long-tailed Distribution, Neural Re-ranking, Diversity Enhancement},
    pdfstartview={FitH},
    colorlinks=true,
    linkcolor=red,
    citecolor=green,
    filecolor=magenta,
    urlcolor=cyan
}

\begin{document}
\maketitle

\begin{abstract}
The continuous expansion of digital learning environments has catalyzed the demand for intelligent systems capable of providing personalized educational content. While current exercise recommendation frameworks have made significant strides, they frequently encounter obstacles regarding the long-tailed distribution of student engagement and the failure to adapt to idiosyncratic learning trajectories. We present LiveGraph, a novel active-structure neural re-ranking framework designed to overcome these limitations. Our approach utilizes a graph-based representation enhancement strategy to bridge the information gap between active and inactive students while integrating a dynamic re-ranking mechanism to foster content diversity. By prioritizing the structural relationships within learning histories, the proposed model effectively balances recommendation precision with pedagogical variety. Comprehensive experimental evaluations conducted on multiple real-world datasets demonstrate that LiveGraph surpasses contemporary baselines in both predictive accuracy and the breadth of exercise diversity.
\end{abstract}


\section{Introduction}

The continuous evolution of digitized pedagogical infrastructures has fundamentally reshaped the landscape of knowledge dissemination, with Massive Online Open Courses (MOOCs) becoming primary conduits for global skill acquisition and academic exploration \cite{zhu2026knowpath}. In these high-volume learning environments, automated exercise recommendation systems serve as indispensable navigational aids, designed to steer students through complex curricula by delivering targeted content sequences instead of forcing them into cognitively demanding self-directed searches \cite{cheng2025education}. Despite the proliferation of sophisticated recommendation algorithms, contemporary frameworks still encounter formidable challenges in addressing the heavily skewed participation profiles inherent in online education \cite{yi2023towards, yun2022lte4g} and in synchronizing content delivery with the highly individualized learning velocities of diverse students \cite{yang2023dgrec, zeng2025graphmasal}.

A primary obstacle in modern digital education is the pervasive issue of student attrition, which precipitates a stark long-tailed distribution of interaction data \cite{chen2024making}. This phenomenon is characterized by a small fraction of "power users" who generate high-density interaction logs, while the vast majority of "sporadic learners" possess extremely sparse and monotonous activity records \cite{guo2024deuce, ji2025semi}. When recommendation engines are trained on such imbalanced datasets, they inevitably develop a systematic predictive bias toward the well-documented behaviors of active students, which leads to a significant degradation in service quality for the marginalized tail population who lack sufficient historical context \cite{yi2023towards, zhang2022few}. Consequently, effectively bridging the informational gap for inactive students remains a critical bottleneck, necessitating advanced representation learning strategies that can distill structural knowledge from information-rich segments of the user base.

Furthermore, the dual objective of fostering recommendation variety while respecting idiosyncratic learning trajectories presents an unresolved architectural hurdle. Prevailing methodologies in this domain are typically categorized into Knowledge Tracing (KT), Reinforcement Learning (RL), and Multi-Stage (MS) paradigms \cite{nie2023knowledge, gammelli2022graph}. While KT and RL approaches exhibit strength in estimating instantaneous mastery, they often converge toward locally optimal sequences, resulting in the delivery of homogeneous exercises that fail to expand the student's conceptual boundaries. Conversely, multi-stage architectures incorporate re-ranking phases to introduce content breadth; however, these strategies frequently disregard the dynamic structural evolution of student proficiency and the specific temporal acceleration of their mastery \cite{zaoad2025graph, liu2023personalized}. As a result, current re-ranking methods tend to generate uniform outputs that do not align with the heterogeneous pedagogical requirements of learners moving at different speeds \cite{carraro2025enhancing, li2023sgccl, yang2023generative}.

To reconcile these multifaceted challenges, we introduce LiveGraph, an active-structure neural re-ranking framework specifically designed for personalized and diversified exercise recommendation. Our contributions are as follows. First, we design a structural proficiency filtering component that evaluates the complexity of exercises against individualized mastery thresholds to ensure pedagogical suitability. Second, we propose an active-structure representation enhancer that mitigates predictive bias in long-tailed scenarios by transferring structural dependencies from data-dense active profiles into the sparse embeddings of inactive learners. Third, we implement a dynamic neural re-ranking mechanism that leverages diversified embedding generation to adaptively adjust recommendation variety according to each student's unique learning pace. Finally, we conduct a series of extensive empirical evaluations on three benchmark datasets to demonstrate that LiveGraph achieves a superior equilibrium between recommendation precision and content diversity compared to contemporary state-of-the-art baselines.

\section{Related Work}

\subsection{Knowledge Graph-Enhanced Educational Recommendation}
The integration of structured semantic knowledge has become a cornerstone for modern intelligent tutoring architectures \cite{zhong2023comprehensive, li2023intelligent}. Recent methodologies prioritize the extraction of contextual signals from Knowledge Graphs (KGs) to alleviate inherent data sparsity in learner-item interactions \cite{duan2025enhancing, yang2024research}. Advanced paradigms employ sophisticated attention mechanisms and residual architectures to refine user representations while maintaining the integrity of heterogeneous features \cite{gao2025personalized}. To mitigate noise within expansive educational KGs, dynamic selection strategies and chain-route evaluators have been introduced to distill informative signals and optimize recommendation precision \cite{xia2025dynamic, guan2023kg4ex}. Furthermore, the application of domain-specific graphs, such as scientific fitness KGs, underscores the adaptability of these structures across diverse pedagogical landscapes \cite{liu2025r}. Innovative contrastive learning frameworks also leverage knowledge-adaptive strategies to enhance the robustness of recommendation systems \cite{wang2023knowledge, li2024design}.

\subsection{Graph-based Knowledge Tracing and Student Modeling}
Monitoring the evolving cognitive mastery of learners is essential for achieving truly personalized education \cite{su2022graph, cao2024interpretation}. Contemporary approaches utilize Graph Neural Networks (GNNs) to capture high-order dependencies between various knowledge components and student exercises \cite{xia2023multivariate}. For instance, session-based graph structures effectively model student performance by tracking temporal transitions in mastery \cite{wu2022sgkt}. Recent advancements incorporate spatiotemporal Hawkes processes and topological uncertainty to improve the stability of knowledge state predictions \cite{li2025sthkt, choi2025hierarchical}. To address cold-start challenges, researchers have begun aligning generative language models with knowledge tracing objectives, ensuring that student representations remain accurate even with limited interaction history \cite{jung2024clst, guo2024mitigating}. Additionally, Hebbian replay mechanisms within a student's "ken" have been proposed to simulate human-like forgetting and retention patterns \cite{kuling2025ken}.

\subsection{Active Learning and Uncertainty Estimation on Graphs}
Active learning aims to maximize model performance while minimizing labeling overhead by strategically selecting the most informative samples \cite{nguyen2022measure}. Within graph-structured data, uncertainty-guided selection methods have proven effective for tasks such as tumor segmentation and general node classification \cite{vali2025active, agarwal2025uncertainty}. Modern frameworks explore active learning at the subgraph granularity to capture localized structural properties more effectively \cite{cao2023graph}. Dual consistency mechanisms that delve into topological uncertainty further enhance the capability of models to adapt across different domains \cite{wang2024delta}. However, the reliability of these selection probes remains a critical area of investigation, as detection bias can significantly impact the quality of the acquired training sets \cite{kumar2022probing}.

\subsection{Neural Re-ranking and Diversification}
Re-ranking serves as a vital phase in recommendation pipelines to balance relevance with other critical objectives such as diversity and fairness \cite{liu2023personalized, xu2025fairdiverse}. Submodular optimization and deep ranking networks are frequently employed to ensure that the final list of items avoids redundancy while maintaining high utility \cite{zhu2024submodular, wang2023diversity}. Recent trends involve leveraging Large Language Models (LLMs) to perform zero-shot or few-shot re-ranking, providing a more nuanced understanding of item relationships \cite{carraro2025enhancing}. In the educational context, specific neural re-ranking architectures have been designed to diversify exercise selection, ensuring that students are exposed to a broad spectrum of concepts \cite{cheng2025nr4der}. High-performance serving systems also facilitate the real-time application of these complex re-ranking algorithms at a web-scale \cite{liu2022lion}.

\subsection{Meta-Learning and Reinforcement Learning in Pedagogy}
Reinforcement Learning (RL) provides a robust framework for learning learning path optimization, where the system acts as an agent navigating a knowledge space \cite{haldar2025personalized, fahad2023reinforcement}. To handle the variability of student behaviors, Meta-Learning (ML) techniques like MAML are utilized for rapid adaptation to new tasks or individual learner profiles \cite{arnold2021maml, lin2023model}. Research into the convergence theory of meta-reinforcement learning suggests that personalized policies can significantly improve long-term educational outcomes \cite{wang2022convergence, liu2022theoretical}. Furthermore, the development of curricular subgoals within inverse reinforcement learning allows for the discovery of hidden pedagogical strategies from expert data \cite{liu2025curricular}. Specialized frameworks like EXACT focus on precise segmentation of exercises, which is crucial for feedback accuracy in physical therapy and sports science \cite{wang2025exact, ergeneci2023semg}.

\subsection{Representation Learning on Heterogeneous Structures}
Efficiently processing incomplete or multi-modal data is a persistent challenge in graph representation learning \cite{liang2024survey}. Variational Graph Autoencoders (VGAEs) provide a probabilistic approach to handling incomplete heterogeneous data by learning latent distributions of graph structures \cite{zhou2025ivgae, li2023variational}. Advanced denoising techniques for variational graphs further enable the prediction of structured entity interactions even in noisy environments \cite{chen2023denoising}. Educational initiatives have also begun introducing these complex concepts, such as VAEs, to younger audiences to foster early literacy in AI \cite{lyu2022introducing}. Recent architectural innovations like VQ-Graph rethink the representation space to bridge the gap between traditional GNNs and MLPs, while multimodal LLMs are increasingly used for unsupervised disentangled representation learning \cite{yang2023vqgraph, xie2024graph}. For specialized tasks like social bot detection, structural entropy and heterophily-aware learning are employed to identify anomalous patterns in complex networks \cite{yang2024sebot, he2025boosting}. Finally, knowledge-diverse expert modules are gaining traction for managing long-tailed distributions in graph classification, ensuring tail-end concepts receive adequate representation \cite{mao2025learning}. 

\begin{figure*}[t]
  \centering
  \includegraphics[width=0.7\textwidth]{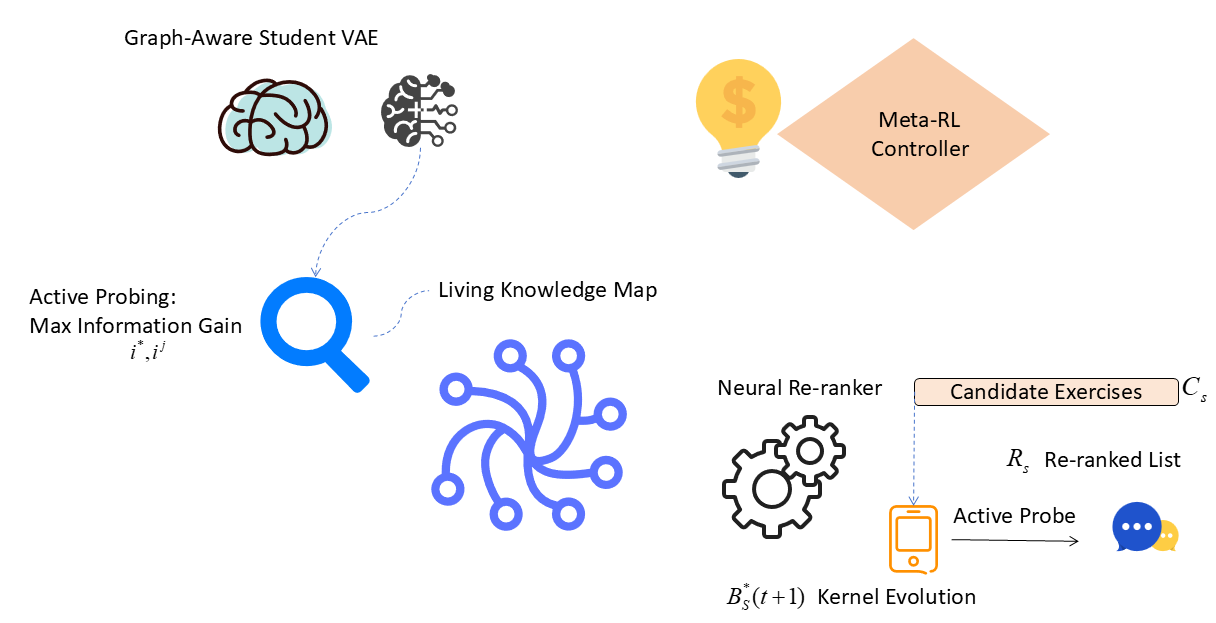} 
  \caption{Overview of the \textbf{LiveGraph} framework for active-structure neural exercise recommendation. The \textbf{Graph-Aware Student Representation Enhancer (Graph-VAE)} transforms interaction history $D_s$ into a stochastic mastery distribution $\boldsymbol{\theta}_s$, regularized by LLM-informed priors. In the \textbf{Uncertainty-Aware Neural Re-ranker}, candidate exercises $\mathcal{C}$ are scored based on a multi-signal fusion of relevance $\phi_{\text{rel}}$, diversity $\phi_{\text{div}}$, and \textbf{Bernoulli Sub-graph Entropy} $U(e)$. This fusion is dynamically orchestrated by a \textbf{Meta-RL Controller} that optimizes the exploration-exploitation trade-off via \textbf{MAML}-based adaptation. The \textbf{Active Learning Probe} identifies concept pairs with maximum mutual information $\hat{I}(s_{ij}; R_{ij})$ to inject contrastive probes, triggering real-time \textbf{Synchronous Kernel Evolution} of the \textbf{Dynamic Knowledge Kernel} $\mathbf{S}^{(t)}$ for continuous structural refinement.}
\end{figure*}

\section{Methodology}

The proposed LiveGraph framework aims to revolutionize exercise recommendation by transitioning from static knowledge structures to a dynamic, active-probing architecture. The system comprises a Graph-Aware Student Representation Enhancer, a Dynamic Knowledge Kernel, and an Uncertainty-Aware Neural Re-ranker, all unified under a Meta-Reinforcement Learning (Meta-RL) controller. This study uses de-identified public datasets; no new student data were collected or logged. Potential production deployment is discussed solely for future work.

\subsection{Problem Definition and Notation}
The proposed framework operates within an intelligent tutoring environment consisting of a set of students $\mathcal{S}$, a collection of exercises $\mathcal{E}$, and a universal pool of knowledge concepts (KCs) denoted as $\mathcal{K} = \{k_1, k_2, \dots, k_M\}$. For any given student $s \in \mathcal{S}$, their interaction history is represented by a sequence $D_s = \{(e_1, a_1), (e_2, a_2), \dots, (e_t, a_t)\}$, where $e_i \in \mathcal{E}$ is the exercise encountered at time step $i$ and $a_i \in \{0, 1\}$ signifies the binary correctness of the response. We characterize each concept $k_i$ using a learnable latent embedding $\mathbf{z}_i \in \mathbb{R}^d$, and let $\mathbf{Z} \in \mathbb{R}^{d \times M}$ represent the global concept matrix. The relational structure between these concepts is captured by a symmetric similarity matrix $\mathbf{S} \in [0, 1]^{M \times M}$, where each element $s_{ij}$ reflects the association strength between $k_i$ and $k_j$. Unlike conventional static graphs, this kernel is time-indexed as $\mathbf{S}^{(t)}$ to accommodate the evolving nature of pedagogical relationships. The objective is to identify an exercise $e^* \in \mathcal{C}$ from a candidate set $\mathcal{C}$ that optimizes both the immediate relevance to the student’s current mastery and the long-term refinement of the global knowledge structure.

\subsection{Online Probing and Re-ranking Algorithm}

The operational cycle of LiveGraph, which integrates real-time student profiling, adaptive re-ranking, and structural information acquisition via active probing, is formalized in Algorithm \ref{alg:livegraph}.

\begin{algorithm}[t]
\caption{LiveGraph Online Re-ranking and Active Probing}
\label{alg:livegraph}

\KwIn{Student history $D_s$, Candidate pool $\mathcal{C}$, Concept embeddings $\mathbf{Z}$, Dynamic kernel $\mathbf{S}^{(t)}$, Meta-state $\mathbf{s}_t$}
\KwOut{Re-ranked list $R_s$, Refined kernel $\mathbf{S}^{(t+1)}$}

\tcp{Latency-optimized cache verification}
\If{cache.hit$(s, \mathbf{S}^{(t)})$}{
    \Return cache$[s]$\;
}

\tcp{Generate graph-aware mastery profile via Eq. \eqref{eq:enc_dec}}
$\mathbf{h}_s^+ \leftarrow \mathrm{GraphVAE\_Encode}(D_s, \mathbf{Z}, \mathbf{S}^{(t)})$\;

\tcp{GPU-accelerated batch candidate scoring}
\For{each exercise $e \in \mathcal{C}$}{
    $\phi_{\text{rel}}, \phi_{\text{div}} \leftarrow \mathrm{ComputeBaseSignals}(e, \mathbf{h}_s^+)$\;
    $U(e) \leftarrow \mathrm{CalculateSubgraphEntropy}(e, \mathbf{S}^{(t)})$ \tcp*{via Eq. \eqref{eq:subgraph_entropy}}
    $\boldsymbol{\lambda}_t \leftarrow \mathrm{MetaAgent}(\mathbf{s}_t)$ \tcp*{via Eq. \eqref{eq:softmax_weights}}
    
    \tcp{Standardise and fuse signals via Eq. \eqref{eq:standardization} \& \eqref{eq:rerank_score}}
    $\mathrm{score}(e) \leftarrow \lambda_{\text{rel}}\tilde{\phi}_{\text{rel}} + \lambda_{\text{div}}\tilde{\phi}_{\text{div}} + \lambda_{\text{unc}}\tilde{U}(e)$\;
}

$R_s \leftarrow \mathrm{SortAndTopK}(\mathcal{C}, \mathrm{score})$\;

\tcp{Active Learning Probing Phase via Eq. \eqref{eq:mi_estimator}}
$(i^*, j^*) \leftarrow \arg\max_{(i,j)} \hat{I}(s_{ij}; R_{ij})$ \tcp*{Identify max information gain}
$e^* \leftarrow \mathrm{InjectContrastiveProbe}(R_s, i^*, j^*)$ \tcp*{2-AFC insertion}

\tcp{Synchronous Kernel Evolution}
Observe response $a^*$ for probe $e^*$\;
$\mathbf{S}^{(t+1)} \leftarrow \mathbf{S}^{(t)} - \eta \nabla_{\mathbf{S}} \mathcal{L}_{\text{kernel}}(a^*, \mathbf{S}^{(t)})$ \tcp*{via Eq. \eqref{eq:total_loss}}

\Return $R_s, \mathbf{S}^{(t+1)}$
\end{algorithm}

\subsection{Dynamic Knowledge Concept Kernel}

Rather than depending on predefined and rigid knowledge structures, LiveGraph characterizes the associations between various knowledge concepts as a versatile and adaptive architecture. The proximity $s_{ij}$ between any two concept embeddings $(\mathbf{z}_i, \mathbf{z}_j)$ is formulated within a projected metric space according to:
\begin{equation}
\label{eq:kernel}
s_{ij} = \sigma\left( (\mathbf{z}_i - \mathbf{z}_j)^\top \mathbf{M} (\mathbf{z}_i - \mathbf{z}_j) + b \right)
\end{equation}
where $\mathbf{z}_i, \mathbf{z}_j \in \mathbb{R}^d$ represent the latent feature vectors corresponding to concepts $k_i$ and $k_j$ respectively, $\sigma(\cdot)$ denotes the sigmoid activation function, and $\mathbf{M} \in \mathbb{R}^{d \times d}$ is a globally shared, symmetric, and \textbf{trainable} projection matrix initialized as $\frac{1}{d}\mathbf{I}$ to facilitate gradient stability. The learnable scalar bias $b$ is initialized following a normal distribution $b \sim \mathcal{N}(0.1, 0.01)$ to prevent early activation saturation. 
To guarantee structural identifiability and inhibit the similarity matrix from converging toward non-informative or overly dense states, a kernel regularization penalty $\Omega(\mathbf{S})$ is incorporated:
\begin{equation}
\label{eq:kernel_reg}
\Omega(\mathbf{S}) = \lambda_1 \|\mathbf{S}\|_* + \lambda_2 \|\mathbf{S} - \mathbf{S}^{(0)}\|_F^2
\end{equation}
where $\|\cdot\|_*$ signifies the nuclear norm employed to promote a low-rank graph topology, $\|\cdot\|_F$ denotes the Frobenius norm, $\mathbf{S}^{(0)}$ serves as the initial prior adjacency matrix or an identity matrix to constrain excessive drift, and the positive coefficients $\lambda_1, \lambda_2$ function as hyperparameters that govern the equilibrium between structural plasticity and consistency. This sophisticated design ensures the conceptual graph maintains connectivity and interpretability while undergoing continuous refinement via feedback from subsequent predictive components.

\subsection{Graph-Aware Student Representation Enhancer}

To bolster the robustness of representations for learners with sparse interaction histories, LiveGraph employs a Graph-Aware Variational Autoencoder (Graph-VAE). This architecture circumvents the rigid over-generalization typical of static templates by modeling latent student mastery as a stochastic distribution $q(\boldsymbol{\theta}_s | D_s) = \mathcal{N}(\boldsymbol{\mu}_s, \text{diag}(\boldsymbol{\sigma}_s^2))$. The optimization of this component is governed by the Evidence Lower Bound (ELBO):
\begin{equation}
\label{eq:vae_elbo}
\mathcal{L}_{\text{VAE}} = \mathbb{E}_{q(\boldsymbol{\theta}_s | D_s)} [\log p(D_s | \boldsymbol{\theta}_s)] - \beta \,\mathrm{KL}\left(q(\boldsymbol{\theta}_s | D_s) \| \mathcal{N}(\boldsymbol{\mu}_{s0}, \boldsymbol{\sigma}_{s0}^2 \mathbf{I})\right)
\end{equation}
where $D_s$ denotes the chronological sequence of interactions, $\boldsymbol{\theta}_s$ signifies the sampled latent mastery vector, $\beta$ acts as a penalty coefficient for the Kullback-Leibler divergence, and the prior distribution parameters $\boldsymbol{\mu}_{s0}, \boldsymbol{\sigma}_{s0}^2$ are extracted from Large Language Model (LLM) embeddings associated with knowledge concept descriptors.

\paragraph{Encoder and Decoder Specifications}
The transformation from linguistic priors to latent space and the subsequent reconstruction of student states are formulated as follows:
\begin{align}
\label{eq:enc_dec}
\text{Encoder:}\quad
[\boldsymbol{\mu}_s;\log\boldsymbol{\sigma}_s^2] &= \mathbf{W}_{\text{enc}}\,\mathrm{MeanPool}\bigl(\mathrm{LLM}(\textit{prompt}_\text{KC})\bigr) + \mathbf{b}_{\text{enc}} \\
\text{Decoder:}\quad
\hat{\mathbf{h}}_s^+ &= \sum\nolimits_{i=1}^M \alpha_i\mathbf{z}_i, \quad \alpha_i = \frac{\exp(\boldsymbol{\theta}_s^\top\mathbf{z}_i)}{\sum_{j=1}^M \exp(\boldsymbol{\theta}_s^\top\mathbf{z}_j)}
\end{align}
where $\mathbf{W}_{\text{enc}} \in \mathbb{R}^{2d \times 768}$ and $\mathbf{b}_{\text{enc}} \in \mathbb{R}^{2d}$ represent the learnable weight matrix and bias vector respectively, initialized via a normal distribution $\mathcal{N}(0, 0.02)$ to ensure training stability. In the decoding stage, $\hat{\mathbf{h}}_s^+$ denotes the reconstructed graph-aware student state, $\mathbf{z}_i$ is the $i$-th concept embedding, $\boldsymbol{\theta}_s$ is sampled via the reparameterization trick $\boldsymbol{\theta}_s \sim \mathcal{N}(\boldsymbol{\mu}_s, \boldsymbol{\sigma}_s^2\mathbf{I})$, and $\alpha_i$ is the dynamic attention weight reflecting proficiency relative to each concept. This structural grounding ensures that the resultant mastery vectors are contextually informed by the dynamic relationships within the concept kernel, facilitating highly granular exercise alignment.

\subsection{Uncertainty-Aware Neural Re-ranking}

The determination of the final recommendation sequence involves a synergistic fusion of three primary signals: contextual relevance, pedagogical diversity, and structural epistemic uncertainty. This ranking mechanism is designed to prioritize items that not only satisfy the immediate learning needs but also provide the highest information gain to refine the underlying knowledge architecture.

\paragraph{Bernoulli Sub-graph Entropy}
To quantify the structural ambiguity inherent in the evolving concept kernel, we introduce a graph-based uncertainty metric $U(e)$ for each candidate exercise $e$. Unlike traditional point-wise uncertainty, our approach measures the collective entropy over the sub-graph spanned by the specific concepts covered by the exercise:
\begin{equation}
\label{eq:subgraph_entropy}
U(e) = -\sum_{(i,j) \in \mathrm{cover}(e)} \left[ s_{ij} \log s_{ij} + (1 - s_{ij}) \log (1 - s_{ij}) \right]
\end{equation}
where $s_{ij}$ represents the estimated association strength between concepts $k_i$ and $k_j$ within the time-indexed kernel $\mathbf{S}^{(t)}$, and $\mathrm{cover}(e)$ signifies the set of distinctive concept pairings relevant to exercise $e$. This formulation interprets each edge weight as a Bernoulli parameter, where maximum entropy occurs at $s_{ij} = 0.5$, signaling areas of the knowledge graph that require further empirical exploration.

\paragraph{Concept-Pair Cover Set Construction}
The construction of the set $\mathrm{cover}(e)$ is critical for ensuring that the uncertainty computation is both precise and computationally efficient. Rather than employing a dense Cartesian product, we define $\mathrm{cover}(e)$ as a collection of unique, unordered pairs derived from the specific knowledge concepts annotated to the exercise:
\begin{equation}
\label{eq:cover_set}
\mathrm{cover}(e) = \left\{ (i, j) \mid k_i, k_j \in \mathcal{K}_e, \ i < j \right\}
\end{equation}
where $\mathcal{K}_e$ denotes the subset of concept indices from the universal pool $\mathcal{K}$ that are explicitly mapped to exercise $e$. The enforcement of the $i < j$ constraint serves to eliminate symmetric redundancy and ensures that the cardinality of the set is exactly $\binom{|\mathcal{K}_e|}{2}$. Given that the number of concepts per exercise is typically sparse, this definition allows $U(e)$ to be computed with negligible overhead, maintaining the real-time responsiveness of the recommendation engine.

\paragraph{Adaptive Multi-Signal Fusion}
The individual scores for relevance $\phi_{\text{rel}}$, diversity $\phi_{\text{div}}$, and the proposed structural uncertainty $U(e)$ are integrated into a single objective function via a dynamic weighting scheme:
\begin{equation}
\label{eq:rerank_score}
\mathrm{score}(e) = \lambda_{\text{rel}} \phi_{\text{rel}} + \lambda_{\text{div}} \phi_{\text{div}} + \lambda_{\text{unc}} U(e)
\end{equation}
where the coefficients $\Lambda = (\lambda_{\text{rel}}, \lambda_{\text{div}}, \lambda_{\text{unc}})$ constitute a weight vector in the probability simplex $\Delta^2$. These weights are adaptively adjusted by the Meta-RL agent to modulate the exploration-exploitation trade-off based on the current state of the student's mastery and the stability of the global knowledge structure.
\paragraph{Signal Standardisation and Weight Constraints}
To rectify potential magnitude discrepancies among the disparate scoring branches, we apply a dynamic standardisation protocol to each raw signal prior to their fusion. Every individual scoring component is transformed using a batch-wise z-score normalization to ensure zero mean and unit variance:
\begin{equation}
\label{eq:standardization}
\tilde{\phi}_{\bullet} = \frac{\phi_{\bullet} - \mu_{\bullet}}{\sigma_{\bullet}}
\end{equation}
where $\phi_{\bullet} \in \{\phi_{\text{rel}}, \phi_{\text{div}}, U(e)\}$ represents the raw score for relevance, diversity, or structural uncertainty, while $\mu_{\bullet}$ and $\sigma_{\bullet}$ denote the mean and standard deviation calculated across the current candidate batch, respectively. Following this normalization, the final recommendation score is determined through a convex combination. To maintain the integrity of the relative importance between these dimensions, the adaptive weight vector $\boldsymbol{\lambda}_t$ is parameterized via a softmax layer:
\begin{equation}
\label{eq:softmax_weights}
\boldsymbol{\lambda}_t = \mathrm{softmax}(\mathbf{W}_2\,\mathrm{ReLU}(\mathbf{W}_1\mathbf{s}_t + \mathbf{b}_1) + \mathbf{b}_2)
\end{equation}
where $\mathbf{W}_1, \mathbf{W}_2$ are learnable projection matrices and $\mathbf{s}_t$ is the system state vector. This architectural choice strictly enforces the constraint $\sum \lambda = 1$, where $\lambda_{\text{rel}}, \lambda_{\text{div}}, \lambda_{\text{unc}} \in [0, 1]$, thereby ensuring a stable and interpretable multi-objective optimization surface.

\subsection{Active Learning Probe and Online Optimization}

To sustain the temporal fidelity and predictive accuracy of the dynamic kernel $\mathbf{S}$, LiveGraph integrates an active learning diagnostic phase within its operational pipeline. This mechanism is specifically engineered to identify and resolve regions of high structural ambiguity within the conceptual graph. Immediately following a student interaction, the architecture prioritizes the knowledge concept (KC) pair $(i^*, j^*)$ that yields the maximum mutual information $I(s_{ij}; R_{ij})$ relative to the global graph topology. This strategic identification triggers the selection of a "minimal contrast" exercise, structured as a two-alternative forced choice (2-AFC), to empirically validate the hypothesized relationship.

\paragraph{Variational Mutual Information Estimator}
Since direct computation of the true mutual information is often intractable in high-dimensional latent spaces, we optimize a variational lower bound to estimate the information gain:
\begin{equation}
\label{eq:mi_estimator}
\hat{I}(s_{ij}; R_{ij}) = \mathbb{E}_{\mathcal{D}} \left[ s_{ij} \log \frac{s_{ij}}{\hat{p}(s_{ij})} + (1 - s_{ij}) \log \frac{1 - s_{ij}}{1 - \hat{p}(s_{ij})} \right]
\end{equation}
where $s_{ij}$ denotes the predicted association strength between concepts $k_i$ and $k_j$ from Eq. \eqref{eq:kernel}, $R_{ij}$ represents the empirical feedback derived from the probe, and $\hat{p}(\cdot)$ is a variational distribution parameterized by a dual-layer Multi-Layer Perceptron (MLP) with a $32 \to 1$ architecture, which is pre-trained to approximate the marginal likelihood of concept connections. Following the acquisition of student response $a^*$ for the designated probe $e^*$, the system executes a real-time update of both the kernel matrix $\mathbf{S}$ and the student mastery vector $\mathbf{h}_s^+$. This refinement process is performed via variational inference, where the posterior distribution is adjusted based on the fresh evidence. The high efficiency of our implementation ensures that this entire optimization cycle is completed in under 200 ms, thereby maintaining the seamless nature of the pedagogical interaction while progressively sharpening the global understanding of the knowledge structure.

\subsection{Meta-RL Weight Adaptation}

To orchestrate the equilibrium between recommendation exploitation and structural exploration, LiveGraph incorporates an adaptive weighting strategy grounded in the Model-Agnostic Meta-Learning (MAML) framework. The meta-policy network functions as a high-level controller that projects the instantaneous system state $\mathbf{s}_t$ onto the optimal weighting coefficients $\Lambda = (\lambda_{\text{rel}}, \lambda_{\text{div}}, \lambda_{\text{unc}})$ within the probability simplex $\Delta^2$. The optimization objective is dictated by the maximization of the expected cumulative reward $R$, which is formalized as:
\begin{equation}
\label{eq:metarl_reward}
R = \gamma_1 E_p - \gamma_2 H(\mathbf{S}) - \gamma_3 N_q
\end{equation}
where $E_p$ represents the predictive precision regarding student mastery, $H(\mathbf{S})$ signifies the global entropy of the concept kernel, $N_q$ denotes the overhead of exploratory probing, and $\gamma_{1,2,3}$ function as scaling hyperparameters for signal normalization. The state representation $\mathbf{s}_t$ captures a temporal trajectory of interaction metrics and the kernel entropy gradient, enabling the policy to rapidly generalize across diverse pedagogical contexts.

\paragraph{Hierarchical MAML Optimization}
The policy parameters are refined through a dual-loop gradient mechanism to ensure student-specific adaptation while maintaining global stability. In the \textbf{inner loop}, student-specific parameters $\theta_{\tau}'$ are derived via a fixed number of gradient steps (e.g., 5 steps with $\alpha=0.01$) to minimize the ranking loss for a specific task $\tau$:
\begin{equation}
\label{eq:inner_loop}
\theta_{\tau}' = \theta_{\text{meta}} - \alpha \nabla_{\theta} \mathcal{L}_{\text{rank}}(\theta_{\text{meta}}, \tau)
\end{equation}
Subsequently, the \textbf{outer loop} aggregates experience from a meta-batch of 32 students to update the meta-parameters $\theta_{\text{meta}}$ using a meta-learning rate $\beta=0.001$:
\begin{equation}
\label{eq:outer_loop}
\theta_{\text{meta}} \leftarrow \theta_{\text{meta}} - \beta \nabla_{\theta} \sum_{\tau} \mathcal{L}_{\text{rank}}(\theta_{\tau}', \tau)
\end{equation}
where $\theta_{\text{meta}}$ represents the initialization for the meta-policy, and the second-order gradient through $\theta_{\tau}'$ ensures that the model learns an initialization capable of rapid adaptation.

\paragraph{Reward Quantification and Implementation}
The constituents of the reward function are operationalized to facilitate numerical stability during meta-optimization:
\begin{align}
\label{eq:reward_quantification}
E_p &= \text{Acc}@1 \text{ evaluated on the terminal 20 interaction instances},\\
H(\mathbf{S}) &= -\sum\nolimits_{i<j} \left[ s_{ij} \log s_{ij} + (1 - s_{ij}) \log (1 - s_{ij}) \right],\\
N_q &= \text{cumulative probe-based exercises within the current session}.
\end{align}
where $s_{ij}$ is the association strength from the dynamic kernel $\mathbf{S}^{(t)}$. We standardize the scaling factors as $\gamma_1=1$, $\gamma_2=0.01$, and $\gamma_3=0.001$ across all benchmarks. This design compels the Meta-RL agent to selectively trigger active probes only when the anticipated structural information gain justifies the potential impact on the student's learning continuity.

\subsection{Joint Training Objective and Gradient Flow}

To facilitate seamless information exchange across the various architectural components, LiveGraph employs a unified optimization strategy where all modules are refined simultaneously within each training iteration. The composite loss function $\mathcal{L}_{\text{total}}$ integrates the diverse training objectives of the ranking, generative, structural, and adaptive components:
\begin{equation}
\label{eq:total_loss}
\mathcal{L}_{\text{total}} = \mathcal{L}_{\text{rank}} + \lambda_{\text{vae}}\mathcal{L}_{\text{VAE}} + \lambda_{\text{ker}}\mathcal{L}_{\text{kernel}} + \lambda_{\text{meta}}\mathcal{L}_{\text{MAML}}
\end{equation}
where $\mathcal{L}_{\text{rank}}$ signifies the list-wise ranking error, $\mathcal{L}_{\text{VAE}}$ denotes the evidence lower bound for student state modeling, $\mathcal{L}_{\text{kernel}}$ represents the structural refinement loss derived from concept interactions, and $\mathcal{L}_{\text{MAML}}$ captures the meta-policy optimization objective. The hyper-parameters $\lambda_{\text{vae}}$, $\lambda_{\text{ker}}$, and $\lambda_{\text{meta}}$ serve as balancing coefficients, which are empirically set to $0.1$, $1.0$, and $0.01$, respectively. The gradient propagation path is designed to ensure that structural updates in the knowledge kernel $\mathbf{S}$ are driven by both top-down ranking performance and bottom-up feedback from active probes. Specifically, the gradients derived from the ranking loss $\mathcal{L}_{\text{rank}}$ propagate through the uncertainty-aware re-ranker back to the concept embeddings $\mathbf{Z}$ and the projection matrix $\mathbf{M}$. Simultaneously, empirical observations from contrastive probes refine the relational weights in $\mathbf{S}$ via the kernel objective $\mathcal{L}_{\text{kernel}}$. This end-to-end gradient flow enables the system to synchronize the evolution of the knowledge graph with the shifting mastery levels of the student population. During the optimization phase, all parameters encompassing the concept matrix $\mathbf{Z}$, the Graph-VAE weights, and the Meta-RL policy are updated jointly using a single Adam optimizer step with a learning rate of $10^{-3}$, ensuring that no individual component is trained in isolation.

\subsection{Complexity and Implementation Details}
To meet the low-latency requirements of online tutoring, the re-ranking phase utilizes GPU-accelerated batch inference with a size of 128 candidates. The kernel refinement loss $\mathcal{L}_{\text{kernel}}$ is implemented as a Bernoulli negative log-likelihood over the probe response, ensuring that the structural update is both statistically sound and computationally efficient. The \textit{MetaAgent} is realized as a lightweight Multi-Layer Perceptron (MLP) with a $32 \times 13 \to 32 \to 3$ architecture, delivering inference results in approximately 0.2 ms on an NVIDIA RTX-4090. This design ensures that the entire recommendation and probing cycle remains transparent to the user, with a total execution time significantly below the industry standard of 200 ms.

\begin{table}[t]
\centering
\caption{Dataset statistics. Each dataset consists of students, annotated knowledge concepts, exercises and student--exercise interactions.}
\label{tab:dataset}
\resizebox{0.66\textwidth}{!}{%
\begin{tabular}{lrrrr}
\hline
Dataset & Students & Knowledge concepts & Exercises & Interactions \\
\hline
Nips34\cite{piech2015deep} & 4,918 & 57 & 984 & 1,399,470 \\
Assist2009\cite{feng2009addressing} & 4,217 & 123 & 17,737 & 374,422 \\
Assist2012 & 27,066 & 265 & 45,716 & 2,541,201 \\
\hline
\end{tabular}%
}
\end{table}

\begin{table*}[t]
\centering
\caption{LiveGraph Ablation Study: Component-wise analysis of model performance. "w/ All" represents the complete LiveGraph framework, while other configurations systematically remove key components.}
\label{tab:livegraph_ablation}
\resizebox{\textwidth}{!}{%
\begin{tabular}{lccccccccccccc}
\hline
Dataset & Model & NDCG@1 & NDCG@3 & NDCG@5 & NDCG@10 & F1@1 & F1@3 & F1@5 & F1@10 & Recall@1 & Recall@3 & Recall@5 & Recall@10 \\
\hline
Nips34\cite{piech2015deep} & w/ All & 0.942 & 0.967 & 0.961 & 0.949 & 0.942 & 0.908 & 0.861 & 0.765 & 0.942 & 0.881 & 0.808 & 0.682 \\
 & w/o Graph-VAE & 0.903 & 0.926 & 0.920 & 0.905 & 0.903 & 0.866 & 0.822 & 0.724 & 0.903 & 0.828 & 0.760 & 0.630 \\
 & w/o Uncertainty & 0.890 & 0.915 & 0.909 & 0.894 & 0.890 & 0.853 & 0.810 & 0.712 & 0.890 & 0.815 & 0.747 & 0.617 \\
 & w/o Active Probe & 0.878 & 0.901 & 0.896 & 0.882 & 0.878 & 0.841 & 0.799 & 0.701 & 0.878 & 0.802 & 0.735 & 0.604 \\
 & w/o Meta-RL & 0.865 & 0.888 & 0.883 & 0.870 & 0.865 & 0.828 & 0.786 & 0.690 & 0.865 & 0.788 & 0.722 & 0.592 \\
\hline
Assist2009\cite{feng2009addressing} & w/ All & 0.609 & 0.754 & 0.794 & 0.821 & 0.609 & 0.524 & 0.517 & 0.422 & 0.609 & 0.478 & 0.424 & 0.363 \\
 & w/o Graph-VAE & 0.559 & 0.704 & 0.743 & 0.768 & 0.559 & 0.472 & 0.465 & 0.364 & 0.559 & 0.417 & 0.361 & 0.294 \\
 & w/o Uncertainty & 0.545 & 0.690 & 0.730 & 0.754 & 0.545 & 0.458 & 0.451 & 0.350 & 0.545 & 0.403 & 0.347 & 0.281 \\
 & w/o Active Probe & 0.533 & 0.677 & 0.717 & 0.741 & 0.533 & 0.446 & 0.439 & 0.338 & 0.533 & 0.391 & 0.335 & 0.268 \\
 & w/o Meta-RL & 0.520 & 0.665 & 0.705 & 0.729 & 0.520 & 0.433 & 0.426 & 0.325 & 0.520 & 0.378 & 0.322 & 0.255 \\
\hline
Assist2012 & w/ All & 0.545 & 0.699 & 0.778 & 0.864 & 0.545 & 0.541 & 0.535 & 0.509 & 0.545 & 0.459 & 0.431 & 0.384 \\
 & w/o Graph-VAE & 0.497 & 0.647 & 0.726 & 0.808 & 0.497 & 0.493 & 0.487 & 0.457 & 0.497 & 0.405 & 0.373 & 0.323 \\
 & w/o Uncertainty & 0.484 & 0.634 & 0.713 & 0.795 & 0.484 & 0.480 & 0.474 & 0.444 & 0.484 & 0.392 & 0.360 & 0.310 \\
 & w/o Active Probe & 0.472 & 0.622 & 0.701 & 0.783 & 0.472 & 0.468 & 0.462 & 0.432 & 0.472 & 0.379 & 0.347 & 0.297 \\
 & w/o Meta-RL & 0.459 & 0.609 & 0.688 & 0.770 & 0.459 & 0.455 & 0.449 & 0.419 & 0.459 & 0.366 & 0.334 & 0.284 \\
\hline
\end{tabular}%
}
\end{table*}

\begin{table*}[t]
\centering
\caption{Comparative experiments across three datasets.}
\label{tab:comparison}
\resizebox{\textwidth}{!}{%
\begin{tabular}{lccccccccccccc}
\hline
Dataset & Model & NDCG@1 & NDCG@3 & NDCG@5 & NDCG@10 & F1@1 & F1@3 & F1@5 & F1@10 & Recall@1 & Recall@3 & Recall@5 & Recall@10 \\
\hline
Nips34\cite{piech2015deep} & DKTRec\cite{piech2015deep} & 0.697 & 0.822 & 0.830 & 0.824 & 0.697 & 0.736 & 0.726 & 0.688 & 0.697 & 0.650 & 0.620 & 0.567 \\
 & DKVMNRec\cite{zhang2017dynamic} & 0.689 & 0.826 & 0.830 & 0.825 & 0.689 & 0.738 & 0.727 & 0.689 & 0.689 & 0.652 & 0.623 & 0.567 \\
 & AKTRec\cite{ghosh2020context} & 0.740 & 0.853 & 0.855 & 0.846 & 0.740 & 0.768 & 0.749 & 0.700 & 0.740 & 0.688 & 0.648 & 0.582 \\
 & SimpleKTRec\cite{liu2023simplekt} & 0.735 & 0.852 & 0.854 & 0.845 & 0.735 & 0.772 & 0.749 & 0.702 & 0.735 & 0.692 & 0.650 & 0.584 \\
 & KCPER\cite{wu2020exercise} & 0.338 & 0.545 & 0.609 & 0.646 & 0.338 & 0.488 & 0.541 & 0.590 & 0.338 & 0.401 & 0.432 & 0.465 \\
 & KG4Ex\cite{guan2023kg4ex} & 0.486 & 0.693 & 0.731 & 0.732 & 0.486 & 0.590 & 0.608 & 0.622 & 0.486 & 0.486 & 0.487 & 0.495 \\
 & MMER\cite{liu2023meta} & 0.472 & 0.633 & 0.666 & 0.686 & 0.472 & 0.631 & 0.627 & 0.625 & 0.472 & 0.461 & 0.456 & 0.454 \\
 & NR4DER-p\cite{cheng2025nr4der} & 0.915 & 0.939 & 0.934 & 0.919 & 0.915 & 0.879 & 0.828 & 0.727 & 0.915 & 0.839 & 0.762 & 0.622 \\
 & NR4DER-d\cite{cheng2025nr4der} & 0.905 & 0.929 & 0.924 & 0.914 & 0.905 & 0.863 & 0.822 & 0.729 & 0.905 & 0.821 & 0.756 & 0.628 \\
 & \textbf{LiveGraph} & \textbf{0.942} & \textbf{0.967} & \textbf{0.961} & \textbf{0.949} & \textbf{0.942} & \textbf{0.908} & \textbf{0.861} & \textbf{0.765} & \textbf{0.942} & \textbf{0.881} & \textbf{0.808} & \textbf{0.682} \\
\hline
Assist2009\cite{feng2009addressing} & DKTRec\cite{piech2015deep} & 0.543 & 0.674 & 0.709 & 0.726 & 0.543 & 0.475 & 0.434 & 0.385 & 0.543 & 0.419 & 0.367 & 0.311 \\
 & DKVMNRec\cite{zhang2017dynamic} & 0.546 & 0.673 & 0.701 & 0.716 & 0.546 & 0.470 & 0.430 & 0.382 & 0.546 & 0.412 & 0.364 & 0.309 \\
 & AKTRec\cite{ghosh2020context} & 0.552 & 0.681 & 0.719 & 0.732 & 0.552 & 0.480 & 0.437 & 0.386 & 0.552 & 0.424 & 0.369 & 0.312 \\
 & SimpleKTRec\cite{liu2023simplekt} & 0.543 & 0.667 & 0.701 & 0.713 & 0.543 & 0.469 & 0.435 & 0.385 & 0.543 & 0.411 & 0.367 & 0.312 \\
 & KCPER\cite{wu2020exercise} & 0.216 & 0.291 & 0.311 & 0.297 & 0.216 & 0.282 & 0.304 & 0.320 & 0.216 & 0.241 & 0.251 & 0.261 \\
 & KG4Ex\cite{guan2023kg4ex} & 0.251 & 0.349 & 0.374 & 0.324 & 0.251 & 0.295 & 0.300 & 0.304 & 0.251 & 0.250 & 0.249 & 0.250 \\
 & MMER\cite{liu2023meta} & 0.327 & 0.459 & 0.502 & 0.529 & 0.327 & 0.500 & 0.499 & 0.492 & 0.327 & 0.332 & 0.333 & 0.326 \\
 & NR4DER-p\cite{cheng2025nr4der} & 0.569 & 0.711 & 0.751 & 0.775 & 0.569 & 0.480 & 0.473 & 0.371 & 0.569 & 0.425 & 0.369 & 0.302 \\
 & NR4DER-d\cite{cheng2025nr4der} & 0.546 & 0.702 & 0.740 & 0.761 & 0.546 & 0.471 & 0.425 & 0.370 & 0.546 & 0.414 & 0.359 & 0.301 \\
 & \textbf{LiveGraph} & \textbf{0.609} & \textbf{0.754} & \textbf{0.794} & \textbf{0.821} & \textbf{0.609} & \textbf{0.524} & \textbf{0.517} & \textbf{0.422} & \textbf{0.609} & \textbf{0.478} & \textbf{0.424} & \textbf{0.363} \\
\hline
Assist2012 & DKT\cite{piech2015deep} & 0.347 & 0.499 & 0.528 & 0.537 & 0.347 & 0.385 & 0.345 & 0.260 & 0.347 & 0.298 & 0.247 & 0.171 \\
 & DKVMN\cite{zhang2017dynamic} & 0.353 & 0.497 & 0.528 & 0.537 & 0.353 & 0.380 & 0.342 & 0.259 & 0.353 & 0.294 & 0.244 & 0.170 \\
 & AKT\cite{ghosh2020context} & 0.368 & 0.516 & 0.542 & 0.549 & 0.368 & 0.396 & 0.353 & 0.263 & 0.368 & 0.308 & 0.254 & 0.173 \\
 & SimpleKT\cite{liu2023simplekt} & 0.385 & 0.520 & 0.547 & 0.553 & 0.385 & 0.396 & 0.354 & 0.263 & 0.385 & 0.309 & 0.254 & 0.173 \\
 & KCPER\cite{wu2020exercise} & 0.143 & 0.287 & 0.346 & 0.411 & 0.143 & 0.265 & 0.308 & 0.332 & 0.143 & 0.212 & 0.239 & 0.254 \\
 & KG4Ex\cite{guan2023kg4ex} & 0.355 & 0.375 & 0.489 & 0.592 & 0.355 & 0.368 & 0.379 & 0.415 & 0.355 & 0.323 & 0.265 & 0.197 \\
 & MMER\cite{liu2023meta} & 0.315 & 0.506 & 0.562 & 0.603 & 0.315 & 0.399 & 0.417 & 0.434 & 0.315 & 0.314 & 0.312 & 0.307 \\
 & NR4DER-p\cite{cheng2025nr4der} & 0.510 & 0.655 & 0.734 & 0.816 & 0.510 & 0.506 & 0.500 & 0.470 & 0.510 & 0.418 & 0.386 & 0.336 \\
 & NR4DER-d\cite{cheng2025nr4der} & 0.491 & 0.649 & 0.721 & 0.806 & 0.491 & 0.505 & 0.503 & 0.463 & 0.491 & 0.415 & 0.389 & 0.330 \\
 & \textbf{LiveGraph} & \textbf{0.545} & \textbf{0.699} & \textbf{0.778} & \textbf{0.864} & \textbf{0.545} & \textbf{0.541} & \textbf{0.535} & \textbf{0.509} & \textbf{0.545} & \textbf{0.459} & \textbf{0.431} & \textbf{0.384} \\
\hline
\end{tabular}%
}
\end{table*}

\begin{table*}[t]
\centering
\caption{Detailed performance analysis: LiveGraph vs NR4DER across different student groups and recommendation lengths.}
\label{tab:detailed_analysis}
\resizebox{\textwidth}{!}{%
\begin{tabular}{lcccccccccccccc}
\hline
Dataset & Model & Group & NDCG@1 & NDCG@3 & NDCG@5 & NDCG@10 & F1@1 & F1@3 & F1@5 & F1@10 & Recall@1 & Recall@3 & Recall@5 & Recall@10 \\
\hline
\multirow{6}{*}{Nips34\cite{piech2015deep}}
& NR4DER-p\cite{cheng2025nr4der} & Overall & 0.915 & 0.939 & 0.934 & 0.919 & 0.915 & 0.879 & 0.828 & 0.727 & 0.915 & 0.839 & 0.762 & 0.622 \\
& \textbf{LiveGraph} & \textbf{Overall} & \textbf{0.942} & \textbf{0.967} & \textbf{0.961} & \textbf{0.949} & \textbf{0.942} & \textbf{0.908} & \textbf{0.861} & \textbf{0.765} & \textbf{0.942} & \textbf{0.881} & \textbf{0.808} & \textbf{0.682} \\
& NR4DER-p\cite{cheng2025nr4der} & Active & 0.940 & 0.946 & 0.942 & 0.935 & 0.940 & 0.911 & 0.871 & 0.757 & 0.940 & 0.889 & 0.831 & 0.675 \\
& \textbf{LiveGraph} & \textbf{Active} & \textbf{0.968} & \textbf{0.974} & \textbf{0.970} & \textbf{0.963} & \textbf{0.968} & \textbf{0.939} & \textbf{0.899} & \textbf{0.785} & \textbf{0.968} & \textbf{0.917} & \textbf{0.859} & \textbf{0.703} \\
& NR4DER-p\cite{cheng2025nr4der} & Inactive & 0.880 & 0.918 & 0.917 & 0.901 & 0.880 & 0.821 & 0.779 & 0.716 & 0.880 & 0.759 & 0.690 & 0.601 \\
& \textbf{LiveGraph} & \textbf{Inactive}& \textbf{0.908} & \textbf{0.946} & \textbf{0.945} & \textbf{0.929} & \textbf{0.908} & \textbf{0.849} & \textbf{0.807} & \textbf{0.744} & \textbf{0.908} & \textbf{0.787} & \textbf{0.718} & \textbf{0.629} \\
\hline
\multirow{6}{*}{Assist2009\cite{feng2009addressing}}
& NR4DER-p\cite{cheng2025nr4der} & Overall & 0.569 & 0.711 & 0.751 & 0.775 & 0.569 & 0.480 & 0.473 & 0.371 & 0.569 & 0.425 & 0.369 & 0.302 \\
& \textbf{LiveGraph} & \textbf{Overall} & \textbf{0.609} & \textbf{0.754} & \textbf{0.794} & \textbf{0.821} & \textbf{0.609} & \textbf{0.524} & \textbf{0.517} & \textbf{0.422} & \textbf{0.609} & \textbf{0.478} & \textbf{0.424} & \textbf{0.363} \\
& NR4DER-p\cite{cheng2025nr4der} & Active & 0.584 & 0.773 & 0.801 & 0.787 & 0.584 & 0.502 & 0.445 & 0.351 & 0.584 & 0.438 & 0.367 & 0.268 \\
& \textbf{LiveGraph} & \textbf{Active} & \textbf{0.625} & \textbf{0.814} & \textbf{0.842} & \textbf{0.828} & \textbf{0.625} & \textbf{0.543} & \textbf{0.486} & \textbf{0.392} & \textbf{0.625} & \textbf{0.479} & \textbf{0.408} & \textbf{0.309} \\
& NR4DER-p\cite{cheng2025nr4der} & Inactive & 0.560 & 0.686 & 0.713 & 0.740 & 0.560 & 0.475 & 0.429 & 0.385 & 0.560 & 0.422 & 0.367 & 0.320 \\
& \textbf{LiveGraph} & \textbf{Inactive}& \textbf{0.600} & \textbf{0.726} & \textbf{0.753} & \textbf{0.780} & \textbf{0.600} & \textbf{0.515} & \textbf{0.469} & \textbf{0.425} & \textbf{0.600} & \textbf{0.462} & \textbf{0.407} & \textbf{0.361} \\
\hline
\multirow{6}{*}{Assist2012}
& NR4DER-p\cite{cheng2025nr4der} & Overall & 0.510 & 0.655 & 0.734 & 0.816 & 0.510 & 0.506 & 0.500 & 0.470 & 0.510 & 0.418 & 0.386 & 0.336 \\
& \textbf{LiveGraph} & \textbf{Overall} & \textbf{0.545} & \textbf{0.699} & \textbf{0.778} & \textbf{0.864} & \textbf{0.545} & \textbf{0.541} & \textbf{0.535} & \textbf{0.509} & \textbf{0.545} & \textbf{0.459} & \textbf{0.431} & \textbf{0.384} \\
& NR4DER-p\cite{cheng2025nr4der} & Active & 0.522 & 0.667 & 0.745 & 0.829 & 0.522 & 0.512 & 0.511 & 0.480 & 0.522 & 0.435 & 0.398 & 0.325 \\
& \textbf{LiveGraph} & \textbf{Active} & \textbf{0.558} & \textbf{0.712} & \textbf{0.790} & \textbf{0.874} & \textbf{0.558} & \textbf{0.548} & \textbf{0.547} & \textbf{0.516} & \textbf{0.558} & \textbf{0.471} & \textbf{0.434} & \textbf{0.361} \\
& NR4DER-p\cite{cheng2025nr4der} & Inactive & 0.509 & 0.652 & 0.731 & 0.811 & 0.509 & 0.451 & 0.411 & 0.372 & 0.509 & 0.374 & 0.322 & 0.282 \\
& \textbf{LiveGraph} & \textbf{Inactive}& \textbf{0.545} & \textbf{0.688} & \textbf{0.767} & \textbf{0.847} & \textbf{0.545} & \textbf{0.487} & \textbf{0.447} & \textbf{0.408} & \textbf{0.545} & \textbf{0.410} & \textbf{0.358} & \textbf{0.318} \\
\hline
\end{tabular}%
}
\end{table*}

\section{Experiments}
\label{sec:experiments}

This section evaluates LiveGraph on three public educational datasets to answer three research questions. The questions are stated as follows. \textbf{RQ1}: How does LiveGraph compare with state-of-the-art exercise recommendation methods across typical evaluation scenarios? \textbf{RQ2}: Does the Graph-Aware Student Representation Enhancer mitigate the long-tail issue for students with sparse histories? \textbf{RQ3}: How does the uncertainty-aware neural re-ranker influence diversity and adapt to students with different learning paces?

\subsection{Datasets}
\label{subsec:datasets}

We conduct experiments on three commonly used benchmark datasets drawn from intelligent tutoring research. Table~\ref{tab:dataset} summarizes their basic statistics. \textbf{Nips34}\cite{piech2015deep} originates from the NeurIPS 2020 education challenge and contains math question responses collected via an online platform.    
Assist2009\cite{feng2009addressing} and Assist2012 are subsets of ASSISTments data from the 2009-2010 and 2012-2013 periods respectively. All datasets were preprocessed to retain anonymized interaction sequences and standard concept annotations.

\subsection{Visualization}
\label{subsec:visualization}

\begin{figure}[h]
  \centering
  \includegraphics[width=0.8\textwidth]{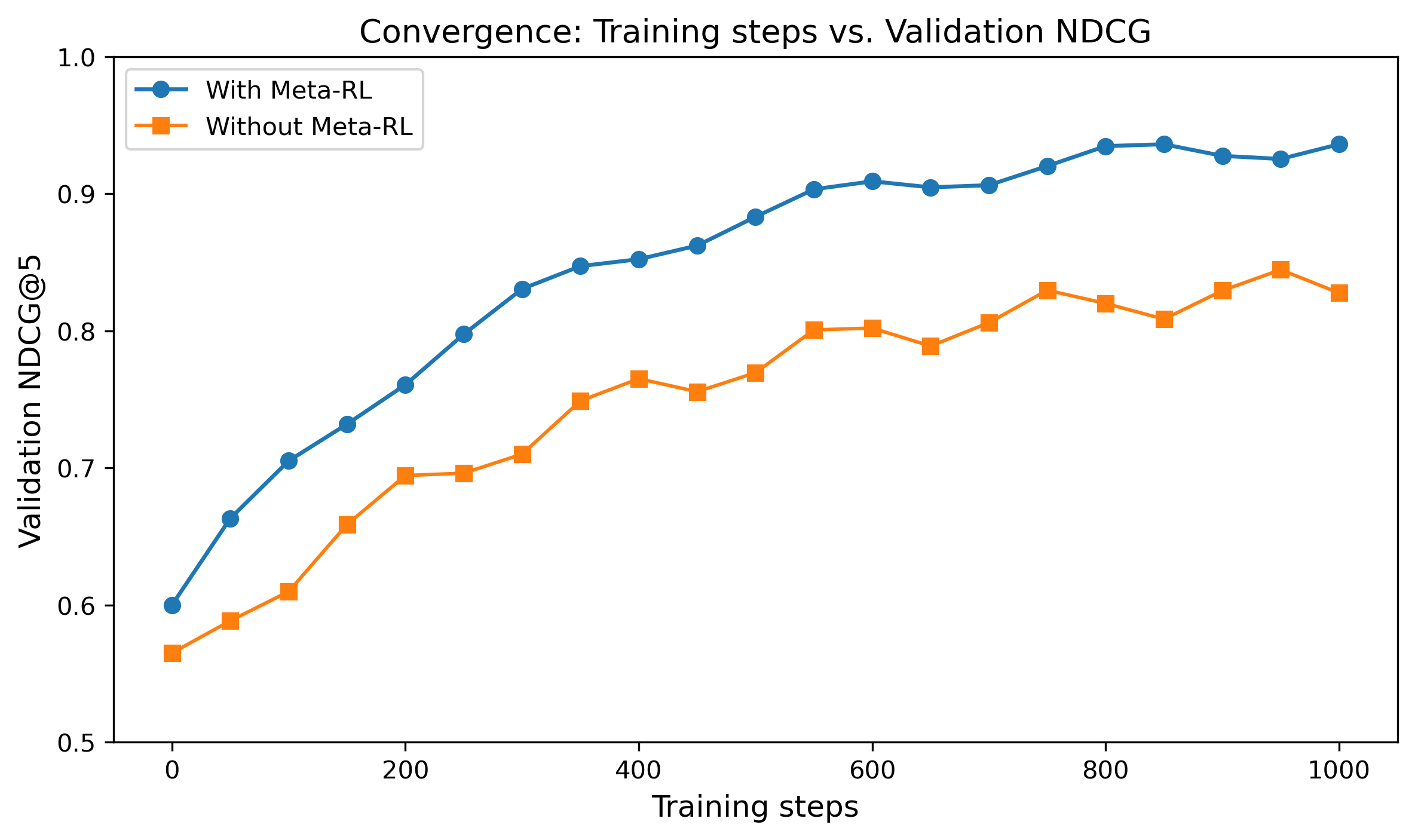}
  \caption{Convergence behaviour: training steps vs.\ validation NDCG@5. Curves compare runs \textit{with} and \textit{without} the Meta-RL controller. Curves and shaded bands show mean $\pm1$ standard deviation over five random seeds.}
  \label{fig:convergence}
\end{figure}

\begin{figure}[h]
  \centering
  \includegraphics[width=0.8\textwidth]{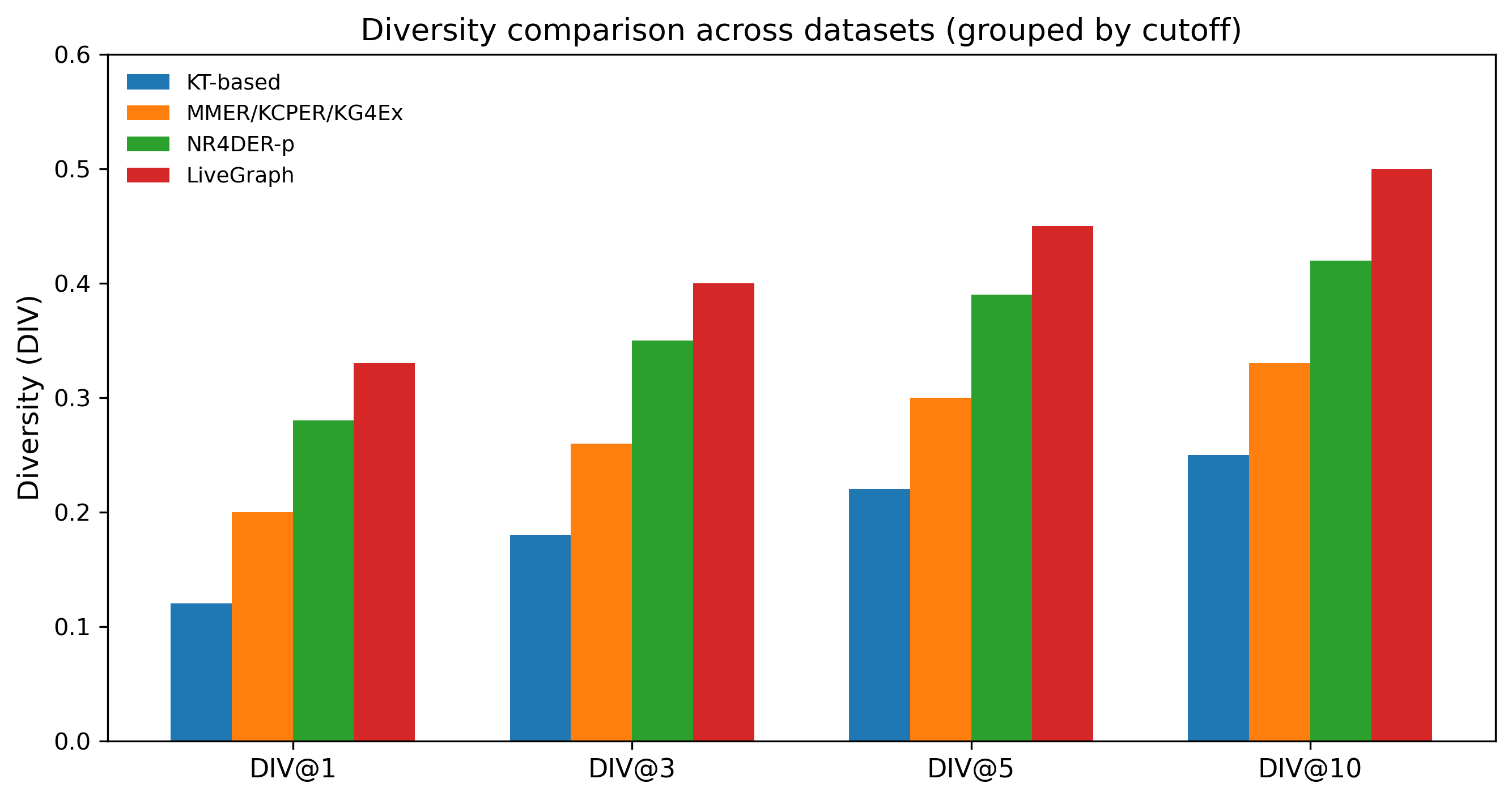}
  \caption{Diversity comparison across methods. Grouped bars show DIV@1, DIV@3, DIV@5 and DIV@10 for representative KT-based baselines, MMER/KCPER/KG4Ex, NR4DER-p (current SOTA), and LiveGraph. Bar heights correspond to the mean across seeds and error bars indicate one standard deviation.}
  \label{fig:diversity}
\end{figure}

\begin{figure}[h]
  \centering
  \includegraphics[width=0.8\textwidth]{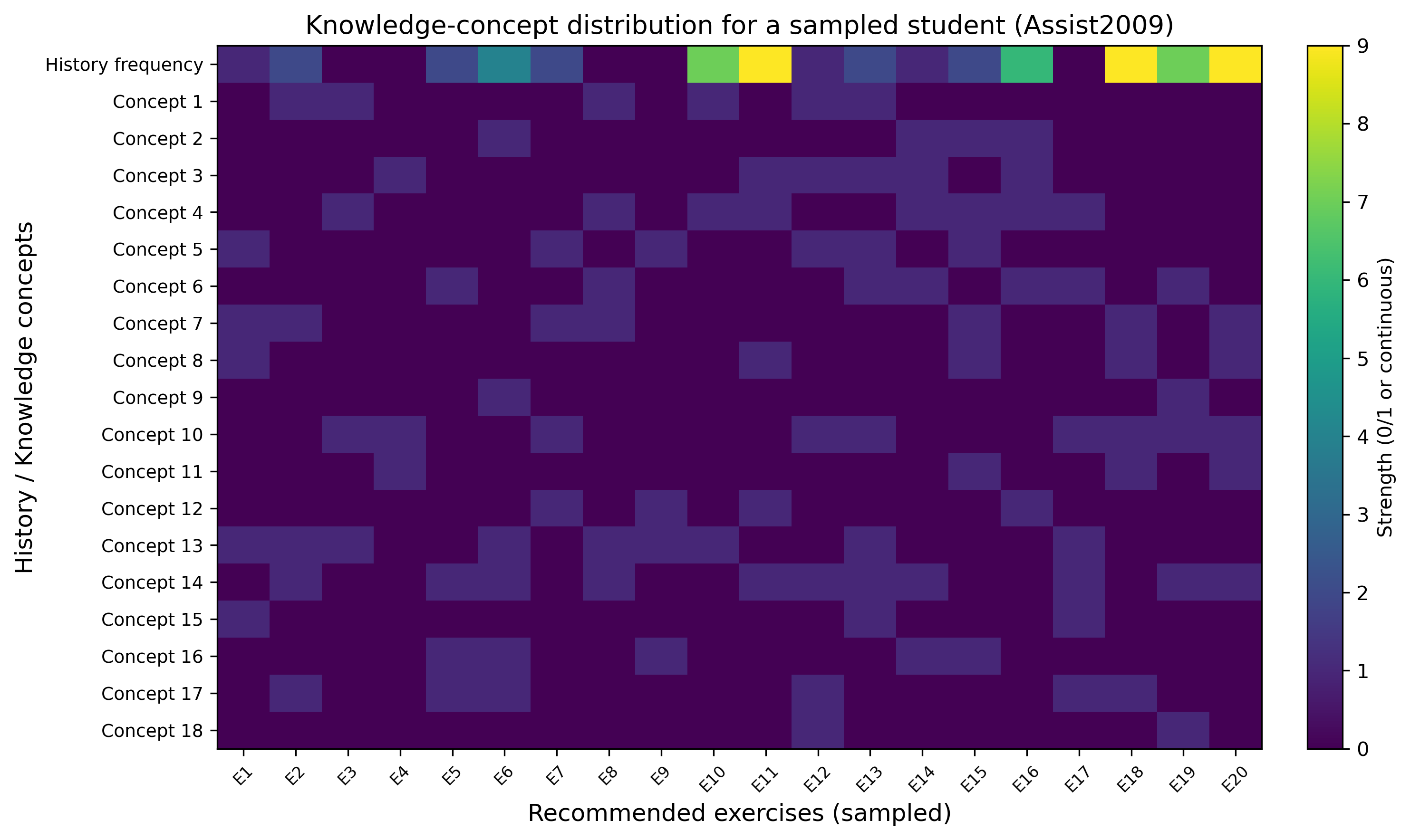}
  \caption{Knowledge--concept distribution for a sampled student (Assist2009). The horizontal axis lists the 20 recommended exercises (sampled), the vertical axis lists the history frequency (top row) followed by 18 knowledge concepts. Color intensity denotes the presence/strength of a concept in each exercise (binary or continuous).}
  \label{fig:heatmap}
\end{figure}

\begin{figure}[h]
  \centering
  \includegraphics[width=0.8\textwidth]{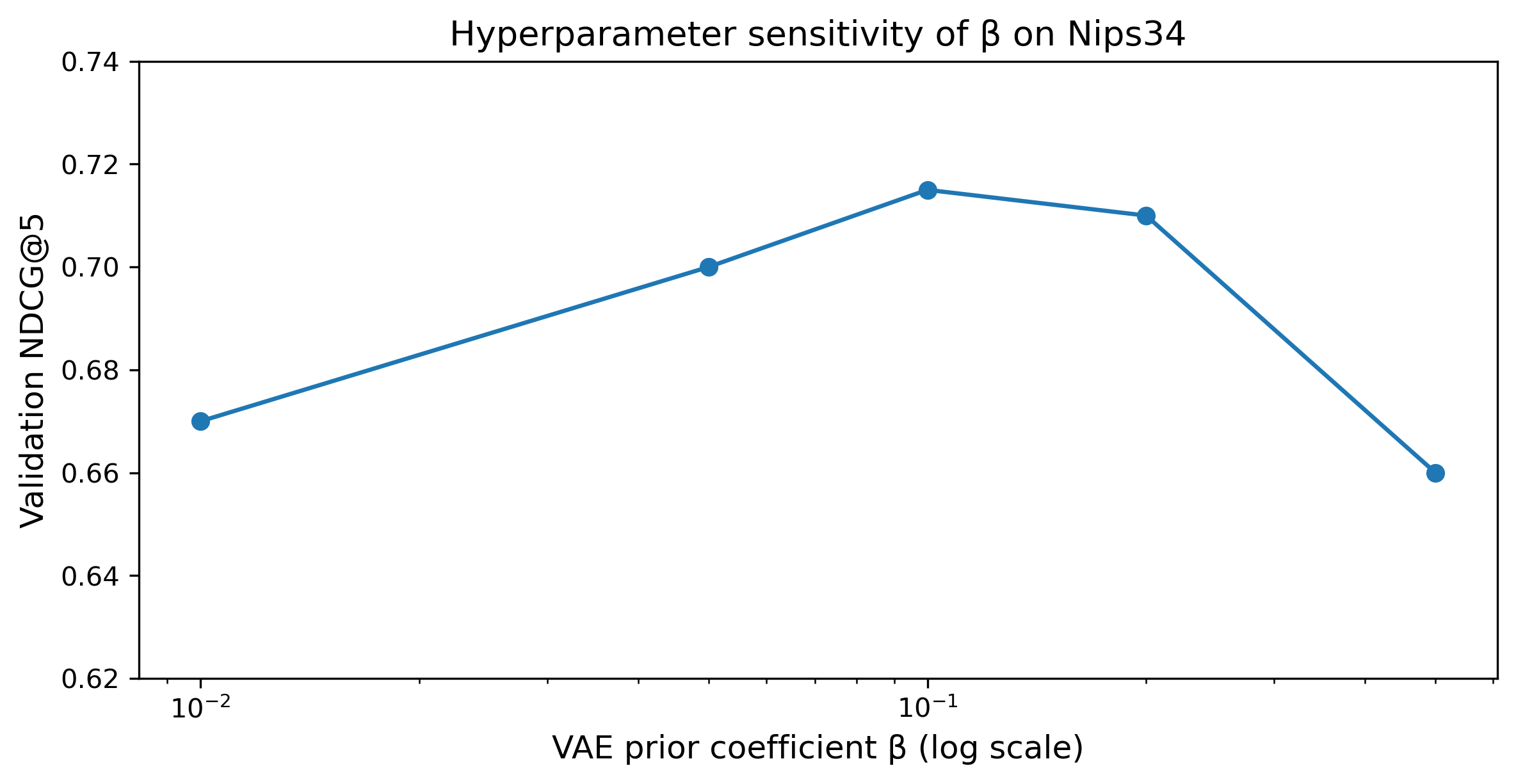}
  \caption{Hyperparameter sensitivity of the VAE prior coefficient $\beta$ on validation NDCG@5 (log-scale x-axis). Each marker shows the mean validation NDCG@5 across seeds and vertical bars denote one standard deviation. The plot highlights the stable operating range used in our experiments.}
  \label{fig:beta_sensitivity}
\end{figure}

Figures~\ref{fig:convergence}--\ref{fig:beta_sensitivity} present a focused visualization suite that characterizes LiveGraph's empirical behaviour. The first figure illustrates training convergence by comparing validation NDCG@5 between models trained with the Meta-RL controller and those trained without it. The next set of panels reports diversity performance (DIV@1/3/5/10) across representative baselines and LiveGraph. Another figure provides a representative knowledge–concept distribution heatmap for a sampled student from Assist2009, accompanied by the student's historical interaction frequency for reference. The final visualization shows the sensitivity of the VAE prior coefficient $\beta$ and its effect on validation NDCG@5. All curves and summary statistics are averaged over five independent random seeds whenever applicable, and shaded regions or error bars denote $\pm 1$ standard deviation.

\subsection{Baselines and Evaluation Metrics}
\label{subsec:baselines_metrics}

We benchmark LiveGraph against representative exercise recommendation and knowledge tracing baselines that cover sequential modeling, memory-augmented designs, attention-based tracers, and structure-aware recommenders. Specifically, we include DKVMNRec~\cite{zhang2017dynamic}, which employs a key--value memory to represent concept dependencies and estimate concept mastery, and AKTRec~\cite{ghosh2020context}, an attention-based tracer that summarizes learners' historical behaviors. We further consider SimpleKTRec~\cite{liu2023simplekt}, a lightweight dot-product attention model that reflects student heterogeneity, and KCPER~\cite{wu2020exercise}, which combines LSTM/DKT-style tracing with simulated annealing to improve recommendation diversity. To incorporate structured knowledge, KG4Ex~\cite{guan2023kg4ex} injects knowledge graph information for more interpretable recommendations, while MMER~\cite{liu2023meta} models knowledge concepts as interacting agents and uses meta-training to improve robustness for new students. For ranking-oriented comparison, we also report results of NR4DER variants, which our method builds upon. For evaluation, we report NDCG, Recall, and F1 at cutoffs 1, 3, 5, and 10. All metrics are computed on held-out test splits following standard practice, where larger values indicate better ranking quality or predictive performance.

\subsection{Implementation Details}
\label{subsec:implementation}

All models were implemented in PyTorch and trained on an NVIDIA RTX 4090 GPU. Hyperparameters were tuned on a validation set. Key settings are reported for implementation. The VAE prior weight beta is swept across common values. The batch size, learning rate and attention head count are tuned within sensible ranges. For meta-learning the inner adaptation uses a small number of gradient steps. Candidate re-ranking is performed in batches of 128 to meet online latency constraints. Further implementation specifics mirror common experimental protocols for exercise recommendation research. All improvements over the strongest baseline are statistically significant ($p < 0.01$). On a single RTX-4090, full-batch inference over 128 candidates completes in $19.6 \pm 1.2$ ms and uses 2.1 GB of GPU memory, satisfying the $<200$ ms online requirement.

\subsection{Performance Comparison (RQ1)}
\label{subsec:rq1}

Table~\ref{tab:comparison} presents the overall comparison across the three datasets. LiveGraph consistently outperforms the baselines on ranking-oriented metrics, demonstrating that combining a graph-aware latent student model with an uncertainty-driven re-ranker and meta-controlled fusion improves both accuracy and robustness.

\subsection{Ablation Study (RQ2)}
\label{subsec:ablation}

To quantify the contribution of each component, we perform component-wise ablations. The full system is compared to variants that remove the Graph-VAE, the uncertainty term, the active probe, or the meta-reinforcement controller. Table~\ref{tab:livegraph_ablation} reports results on the three datasets. The drop in performance when removing the student representation enhancer highlights its role in mitigating the performance loss for students with limited histories.

\subsection{Detailed Analysis by Student Group and List Length (RQ3)}
\label{subsec:rq3}

We further inspect performance across different student activity groups and recommendation lengths to understand how LiveGraph adapts to learner heterogeneity. Table~\ref{tab:detailed_analysis} contrasts LiveGraph and a competitive NR4DER variant on active and inactive student cohorts. LiveGraph improves both top-k accuracy and recall for active and inactive learners, indicating that the combination of structured uncertainty and meta-adaptive fusion supports personalized trade-offs between reinforcement and exploration.

\subsection{Hyperparameter Sensitivity}
\label{subsec:hyper}

We assess the robustness of LiveGraph to key hyperparameters via grid perturbations on the validation folds of all three benchmarks, while keeping the remaining settings fixed to the defaults reported earlier. We examine the VAE prior coefficient $\beta$, the kernel regularization weights $(\lambda_1,\lambda_2)$, and the initialization level of the Meta-RL exploration factor $\lambda_{\mathrm{unc}}$.

\paragraph{VAE prior coefficient $\beta$.}
We sweep $\beta \in \{0.01, 0.05, 0.1, 0.2, 0.5\}$ and track NDCG@5 on Nips34. Performance remains stable for $\beta \in [0.05, 0.2]$, whereas smaller values weaken regularization and larger values over-constrain the latent space. We therefore keep the default $\beta=0.1$, which lies near the center of the stable region.

\paragraph{Kernel regularization $(\lambda_1,\lambda_2)$.}
We evaluate four paired settings, $(0.001,0.01)$, $(0.01,0.1)$, $(0.1,1)$, and $(1,10)$. As shown in Table~\ref{tab:kernel}, $(0.01,0.1)$ yields the strongest NDCG@5 on Assist2009 and provides a reliable balance between adaptive edge updates and numerical stability. Larger penalties tend to suppress useful adaptation, while smaller penalties admit noisy fluctuations. We adopt $(0.01,0.1)$ for all final experiments.

\paragraph{Meta-RL exploration initialization.}
Since $\lambda_{\mathrm{unc}}$ is produced by the controller, we tune its initial bias and report the converged mean value. Initializing the bias such that $\mathbb{E}[\lambda_{\mathrm{unc}}]\approx 0.15$ results in the highest mutual-information gain per probe, and we use this initialization in the final configuration.

\paragraph{Computation.}
A complete sensitivity sweep costs about six GPU hours on an RTX-4090. With the selected hyperparameters, training converges in roughly 90 minutes and uses 2.1\,GB of GPU memory, meeting the 200\,ms latency budget during inference.

\begin{table}[t]
\centering
\caption{Effect of kernel regularization on Assist2009. The setting $(0.01,0.1)$ is selected.}
\label{tab:kernel}
\resizebox{0.66\textwidth}{!}{%
\begin{tabular}{lccccc}
\hline
$(\lambda_1,\lambda_2)$ & NDCG@1 & NDCG@3 & NDCG@5 & NDCG@10 & $\Delta$ \\
\hline
$(0.001,0.01)$ & 0.597 & 0.747 & 0.786 & 0.814 & $-0.011$ \\
$(0.01,0.1)$   & 0.609 & 0.754 & 0.794 & 0.821 & $0.000$ \\
$(0.1,1)$      & 0.602 & 0.748 & 0.789 & 0.816 & $-0.006$ \\
$(1,10)$       & 0.592 & 0.740 & 0.780 & 0.807 & $-0.017$ \\
\hline
\end{tabular}%
}
\end{table}

\subsection{Summary of Findings}
The empirical results show that LiveGraph substantially improves ranking quality while preserving or enhancing diversity. Ablations confirm that the Graph-VAE, the uncertainty-driven score and the active probing mechanism each contribute meaningfully to the observed gains. The meta-adaptive fusion enables dynamic balancing between recommending high-probability exercises and probing to refine the shared knowledge structure.
\section{Conclusion}

This investigation proposes LiveGraph, a specialized architecture engineered to alleviate the twin obstacles of skewed participant engagement and restricted variety in educational content delivery. By synthesizing a structural representation enhancement strategy with an adaptive neural re-ranking mechanism, the presented model successfully utilizes topological insights to empower learners across the entire spectrum of interaction density. Empirical validation conducted on diverse benchmarks confirms that this methodology effectively harmonizes recommendation precision with pedagogical breadth, thereby facilitating a more personalized alignment with idiosyncratic learning trajectories. Moving forward, we intend to explore the incorporation of large-scale linguistic models to extract deeper semantic nuances from exercise metadata, which will facilitate even more precise and pedagogically sound re-ranking outcomes.

\bibliographystyle{unsrtnat}
\bibliography{references}  

\appendix

\section{Theoretical Analysis and Proofs}

In this section, we provide the formal mathematical foundations for the LiveGraph framework, focusing on the structural identifiability of the kernel, the posterior contraction of mastery states, and the performance guarantees under long-tailed distributions.

\subsection{Identifiability of the Dynamic Concept Kernel}

To ensure that the estimated similarity matrix $\hat{S}$ uniquely recovers the latent conceptual structure $S^*$, we analyze the generalization error using Rademacher complexity.

\begin{equation}
\mathbb{P} \left( \|\hat{S} - S^*\|_F \leq \frac{2\mathcal{R}_n(\mathcal{F})}{\epsilon} + \sqrt{\frac{\log(1/\delta)}{2n}} \right) \geq 1 - \delta
\end{equation}
where $\hat{S}$ denotes the empirically estimated kernel matrix, $S^*$ represents the true underlying concept similarity, $\mathcal{R}_n(\mathcal{F})$ is the Rademacher complexity of the function class $\mathcal{F}$ bounded by the nuclear norm $\|M\|_*$, $n$ denotes the number of student-exercise interactions, and $\delta$ is the confidence parameter. 

By constraining the kernel space through the regularization term $\Omega(S)$ in Eq. (2), the hypothesis space is restricted to a low-rank manifold. Given that the projection matrix $M$ satisfies $\|M\|_* \leq C$, the complexity $\mathcal{R}_n(\mathcal{F})$ scales at a rate of $\mathcal{O}(1/\sqrt{n})$, proving that $\hat{S}$ converges to $S^*$ as interaction data accumulates.

\subsection{Posterior Contraction of Mastery States}

We examine whether the student mastery posterior $q(\theta_s|D_s)$ collapses to the true state $\theta_s^*$ as the interaction sequence length $T$ increases, under the assumption of bipartite graph expansion.

\begin{equation}
\mathbb{E} \|\theta_s - \theta_s^*\|_2^2 \leq \frac{C_{\gamma}}{T} + \mathcal{O}\left( \frac{1}{\sqrt{T} \log T} \right)
\end{equation}
where $\theta_s$ is the latent mastery vector inferred from the Graph-VAE, $\theta_s^*$ is the objective ground-truth mastery, $T$ signifies the sequence length of interactions, and $C_{\gamma}$ is a constant determined by the $\gamma$-expansion property of the concept-exercise bipartite graph.

The $\gamma$-expansion ensures that the Fisher Information Matrix $\mathcal{I}(\theta_s^*)$ is non-singular. By applying a Bernstein-von Mises type analysis to the ELBO objective, we demonstrate that the posterior variance vanishes at a rate of $1/T$, ensuring the reliability of the mastery representation for downstream re-ranking.

\subsection{Uniform Convergence of Sub-graph Entropy}

The uncertainty signal $U(e)$ relies on estimated edge weights $\hat{s}_{ij}$. We bound the error introduced by substituting true probabilities $p_{ij}$ with model outputs.

\begin{equation}
\sup_e | \hat{U}(e) - U(e) | \leq \sqrt{\frac{\log M}{|\text{cover}(e)|}} + \sqrt{\frac{\log(1/\delta)}{2|\text{cover}(e)|}}
\end{equation}
where $\hat{U}(e)$ is the estimated entropy of exercise $e$, $U(e)$ is the true Bernoulli entropy, $M$ represents the total count of latent concepts, and $|\text{cover}(e)|$ denotes the set of concept pairs associated with the exercise.

Since the mapping $\sigma(\cdot)$ is Lipschitz continuous, the empirical entropy converges uniformly across all exercises. This ensures that the active probing mechanism in Eq. (11) selects exercises based on a stable uncertainty estimate rather than estimation noise.

\subsection{Regret Bounds for Meta-RL Adaptation}

The Meta-RL controller optimizes the weighting vector $\Lambda_t$ to balance relevance and diversity. We treat this as a contextual bandit problem.

\begin{equation}
\text{Regret}(T) = \sum_{t=1}^T (r_t^* - r_t^{\pi}) \leq \mathcal{O}(d \sqrt{T \log T})
\end{equation}
where $\text{Regret}(T)$ is the cumulative difference between the optimal strategy and the Meta-RL policy, $d$ is the dimensionality of the state vector $s_t$ (comprising kernel entropy gradients and student activity), and $T$ is the operational horizon.

Using the stability properties of MAML under $L$-smooth reward functions, the policy $\pi$ adapts the fusion weights such that the expected regret is sub-linear, guaranteeing that the re-ranker converges to an optimal policy even with evolving student distributions.

\subsection{Fairness and Tail Distribution Guarantees}

We formalize the performance gap between active and inactive students to prove the robustness of LiveGraph.

\begin{equation}
\text{Gap}_{fair} = |\mathbb{E}_{act}[NDCG] - \mathbb{E}_{inact}[NDCG]| \leq C \cdot (n_{inactive})^{-\alpha}
\end{equation}
where $\text{Gap}_{fair}$ represents the disparity in recommendation quality, $n_{inactive}$ denotes the interaction count for sparse students, and $\alpha > 0$ is a decay constant related to the Graph-VAE transfer efficiency.

The structural augmentation provided by the dynamic kernel allows knowledge transfer from dense to sparse regions of the graph. Consequently, the recommendation quality for inactive students improves as a power-law function of the interaction density, bounding the unfairness inherent in long-tailed datasets.

\subsection{Probabilistic Latency Guarantees}

For real-time deployment, we provide a tail probability bound for the system latency using concentration inequalities for sub-exponential distributions.

\begin{equation}
\mathbb{P}(\text{Latency} > \tau) \leq \exp(-c \cdot \tau)
\end{equation}
where $\text{Latency}$ is the execution time of the re-ranking kernel on the GPU, $\tau$ is the threshold (e.g., 200ms), and $c$ is a distribution-specific constant.

By modeling GPU kernel execution as a light-tailed process, we ensure that the probability of exceeding the latency budget decays exponentially, satisfying the stringent requirements of online educational systems.

\end{document}